\documentclass[aip,jcp,amsmath,amssymb,reprint]{revtex4-1}
\usepackage[pdftex]{graphicx}
\usepackage{color}
\usepackage{dcolumn}
\usepackage{bm}

\begin{document}

\preprint{AIP/123-QED}

\title{Study of dynamical heterogeneities in aging colloidal nanoclay suspensions}
\author {Paramesh Gadige} 
\affiliation{Soft Condensed Matter Group, Raman Research Institute, C. V. Raman Avenue, Sadashivanagar, Bangalore 560 080, INDIA}
\author {Debasish Saha}
\affiliation{Soft Condensed Matter Group, Raman Research Institute, C. V. Raman Avenue, Sadashivanagar, Bangalore 560 080, INDIA}
\author {Sanjay Kumar Behera}
\affiliation{Soft Condensed Matter Group, Raman Research Institute, C. V. Raman Avenue, Sadashivanagar, Bangalore 560 080, INDIA}
\author {Ranjini Bandyopadhyay}
\email{ranjini@rri.res.in}
\affiliation{Soft Condensed Matter Group, Raman Research Institute, C. V. Raman Avenue, Sadashivanagar, Bangalore 560 080, INDIA}
\date{\today}

\begin{abstract}
An aqueous suspension of the synthetic clay Laponite can undergo a transition from a liquid-like ergodic state to a glass-like nonergodic arrested state. In an observation that closely resembles the dynamical slowdown observed in supercooled liquids subjected to a rapid temperature quench, the phenomenon of kinetic arrest in Laponite suspensions is accompanied by a growth in the $\alpha$-relaxation or structural relaxation time with increasing sample aging time, $t_{w}$. The ubiquitous dynamic slowdown and fragile behavior observed in glass forming liquids approaching the glass transition is typically ascribed to the heterogeneous dynamics and the growth in the size of distinct dynamical heterogeneities. In this article, we present the characterization of the dynamical heterogeneities in aging colloidal Laponite clay systems by invoking the three-point dynamic susceptibility formalism. The average time-dependent two-point intensity autocorrelation and its sensitivity to the control parameter $ t_{w} $ are probed in dynamic light scattering experiments.  Distributions of relaxation time scales deduced from Kohlrausch-Williams-Watts equation widen with increasing $ t_{w} $ signifying the heterogeneous dynamic slowdown. A suitable formalism to calculate three-point correlation function is employed for aging colloidal suspension where the main control parameter is $t_{w}$. The calculated three-point dynamic susceptibility exhibits a peak, with the peak height increasing with evolving $ t_{w}$. The number of dynamically correlated particles, deduced from the peak-height, is seen to initially increase  with increasing $ t_{w}$ at a fast rate, before eventually slowing down close to the non-ergodic transition point. This observation is in agreement with published reports on supercooled liquids. Our study confirms the presence and growth of dynamical heterogeneities in soft glassy suspensions of Laponite, thereby shedding new light on the fragile supercooled liquid-like dynamics of aging suspensions of these anisotropic, charged, colloidal clay nanoparticles.
%
\end{abstract}

\maketitle
\section{Introduction}
Glasses are characterized by amorphous order and solid-like rigidies and have a wide range of applications in our day-to-day lives \citep{{Wolynes},{Ediger}}. Supercooled liquids and slightly polydispersed colloidal suspensions form glasses as the relevant control parameters,  temperature ($T$) and volume fraction ($ \phi $) respectively, are changed appropriately \citep{{Ediger},{Pusey}}. Their dynamics slow down dramatically with decreasing $T$ or increasing $\phi$ and their viscosities ($ \eta $) or structural relaxation timescales ($ \tau_{\alpha} $) grow by several orders of magnitude. Below the glass transition temperature $ T_{g} $ or above the glass transition volume fraction $ \phi_{g}$  ($\phi_{g}=0.58$ for hard-sphere colloids), $\eta $ becomes so large that no flow or structural relaxation can be detected in the system. Glass-forming liquids are called strong if $\tau_{\alpha} $ (or $ \eta $) shows a nearly Arrhenius growth as a function of the relevant control parameter, whereas they are termed fragile if a super-Arrhenius trend is followed \citep{{Angel},{Angel2}}. In many theories and phenomenological models, the dramatic dynamical slowdown and the fragile behavior of glass-forming liquids are understood by considering the cooperative movement of the constituent molecules or particles of the glass-forming liquid  \citep{{Adam},{Berthier}}. The distinct regions exhibiting correlated particle motion are called Dynamical Heterogeneities (DHs). The sizes of the DHs, or alternatively, the number of molecules or particles showing correlated movement ($ N_{corr} $), grow as the glass transition is approached. The glass transition is manifested by a remarkable slowdown in the dynamics which results in the observed increase in $\tau_{\alpha}$. Using multi-point correlation function techniques, simulations and experiments have established the presence of  DHs  \cite{{Richert}} and the growth of $N_{corr}$ in glass-forming liquids when their temperatures are quenched rapidly \citep{{Ediger2},{Berthier},{Stevenson},{Yodh},{Stillinger},{Dalle},{Dalle2},{Brambilla}}.  Experimentally, such measurements in supercooled liquids are typically performed using non-linear dielectric susceptibility \citep{{Crauste},{Brun},{Berthier2},{Bauer}}. 
  
 The present work characterises the DHs in a fragile colloidal glass. Colloidal clay suspensions are known to exhibit rich phase behavior, showing fluid, gel, ordered and disordered phases \cite{{Herman},{ruzika},{joshi},{ruzika2}}, and serve as excellent model systems to mimic the behaviour of atomic systems \cite{vm_review}. One ubiquitous example of a model colloidal system is Laponite, a synthetic smectite clay which has been studied extensively for its  soft glassy rheology \cite{cates_sgr} and interesting aging properties as it spontaneously transforms from a liquid to a glass with increasing waiting  time (time since preparation), $t_{w}$ \cite{{Mourchid},{bandyopadhyay_prl},{joshi_JCP},{saha_EPL}}. In this article, we characterize the growth of $N_{corr}$ in synthetic Laponite clay suspensions at several $t_{w}$. 
 
 Aqueous suspensions of Laponite clay are known to form Wigner repulsive glasses \cite{{DBonn98},{ruzika},{sahaLagm},bandyopadhyay_prl} in the concentration range $2-3.5$ wt\% due to the buildup of long range inter-particle electrostatic repulsive interactions with increasing $t_{w}$ \cite{Mourchid,bandyopadhyay_prl}.  The  growth of its $\alpha$ relaxation time in a super-Arrhenius manner as $ t_{w} $ increases has been established experimentally \cite{Mourchid,ruzica,saha}. 
 Recent studies in our group have established that the slowdown in the dynamics of Laponite suspensions resembles the slowdown reported in fragile glass-forming molecular liquids if $ t_{w} $ of the former system is mapped to ($1/T$) of the latter in the Arrhenius and the Vogel-Fulcher-Tammann equations, which represent, respectively, the secondary and primary relaxation modes \cite{{saha},{saha2}}. 
In an earlier study, S. Jabbari-Farouji {\it et.al.} \citep{{Jabbari},{Jabbari1}} reported DHs in Laponite suspensions by studying the particles' rotational and translational diffusion coefficients as a function of $ t_{w}$. It was observed that the rotational diffusion of the constituent particles slows down at a faster rate than their translational motion. However, no report was found in the literature  on estimates of the sizes of the DHs in aging Laponite colloidal glasses. 
In this study, we employ the three-point correlation function formalism \citep{{Dalle},{Dalle2},{Brambilla}} to probe DHs and $ N_{corr}$ in Laponite suspensions with changing $ t_{w} $. The relaxation dynamics of Laponite suspensions were studied by analysing the intensity autocorrelation functions obtained in dynamic light scattering experiments as a function of the age of the Laponite suspension, $t_{w}$. The scattering function decay curves are fitted by parametrizing the two-step relaxation equation (showing exponential and stretched exponential decays and representing, respectively, the secondary and primary relaxation processes of the Laponite suspension)  with the control parameter $ t_{w} $. The three-point correlation functions estimated using this analysis, which are the $ t_{w} $ derivatives of the decaying scattering function curves, exhibit peaks whose heights increase with increasing $t_{w}$. $N_{corr}$ is calculated from the peak height data and its evolution with $\tau_{\alpha}/{\tau_{g}}$, where $\tau_{g}$ is the non-ergodic transition time of the Laponite suspension, bears a striking resemblance to exisiting calculations for the growth of correlations in rapidly quenched supercooled liquids. We believe that our study, which demonstrates the existence of remarkable similarities between the kinetic arrest pheonemena in Laponite clay suspensions and in supercooled liquids, provides valuable additional insight into the colloidal glass transition. 
\section{Sample preparation and Experimental methods}
Laponite RD$^{\circledR}$(BYK, Inc.) powder was procured from Southern Clay products. As clay particles are hygroscopic in nature, the powder was dried in a hot oven at 120$^{\circ}$C  for 16 hours. 3.0 wt$\% $ ($C_{L}$) Laponite concentration ($\phi_{L}$ = 1.18$\times$10$^{-2}$) suspension was prepared by adding the dried powder slowly to Milli-Q water (resistivity 18.2 M$\Omega$-cm). $ C_{L} $ was chosen to lie in a region of the phase space where the Laponite suspension is expected to form a Wigner glass \cite{{DBonn98},{ruzika}}. The suspension was stirred vigorously for 1 hour using a magnetic stirrer. The resulting optically clear and homogeneous suspension was filtered using a 0.45 $\mu$m Millipore Millex-HV grade filter using a syringe pump at a constant flow rate of 3 ml/min. A very small volume fraction ($\phi_{PS}$ = 5.66$\times$10$^{-5} $) of  polystyrene (PS) probe particles (100 nm in size) was added and mixed homogeneously. These PS beads are expected to act as light scatterers in the otherwise highly transparent Laponite suspensions \citep{{Cipel}}. The Laponite-PS suspensions were subsequently sealed in a cuvette. The aging time or waiting time, $t_{w}$, was calculated from the time when the stirring of the suspension was stopped and the cuvette was sealed. Auto-correlation functions of the intensity scattered by these suspensions were recorded in dynamic light scattering (DLS) experiments \cite{Bern} using a Brookhaven Instruments Corporation BI-200SM spectrometer and a BI-9000AT digital autocorrelator. A constant temperature of 25$^{\circ}$C was maintained  using a temperature controller (Polyscience Digital) attached to the DLS system. Details of the set-up are given elsewhere \citep{saha}. The normalized intensity autocorrelation function of the scattered light,  $g^{(2)}(q,t) = \frac{<I(q,0)I(q,t)>}{<I(q,0)>^{2}}$, was recorded as a function of delay time $t$. Here, $q$ and $I(q,t)$ are the scattering wave vector and the intensity of the scattered light at a particular $q$ and $t$ respectively. $q$ is related to the scattering angle $\theta$ by the equation $q=(4\pi n/\lambda)\sin(\theta/2)$, where $n$ and $\lambda$ are the refractive index of the medium ($n$ = 1.334 ) and the wavelength of the laser ($ \lambda$ = 532 nm) respectively \citep{{Bern}}. The intensity autocorrelation data was recorded at $\theta$ = 90$^{\circ}$ and $\theta$ = 60$^{\circ}$. The three point correlation functions were computed by taking the derivatives of the time-dependent self-intermediate scattering function with respect to $t_{w}$ using Mathematica.
\section{Results and Discussions}
\subsection{Characterising the structural relaxation process}

\begin{figure}[!t]
\includegraphics[width=3.3in]{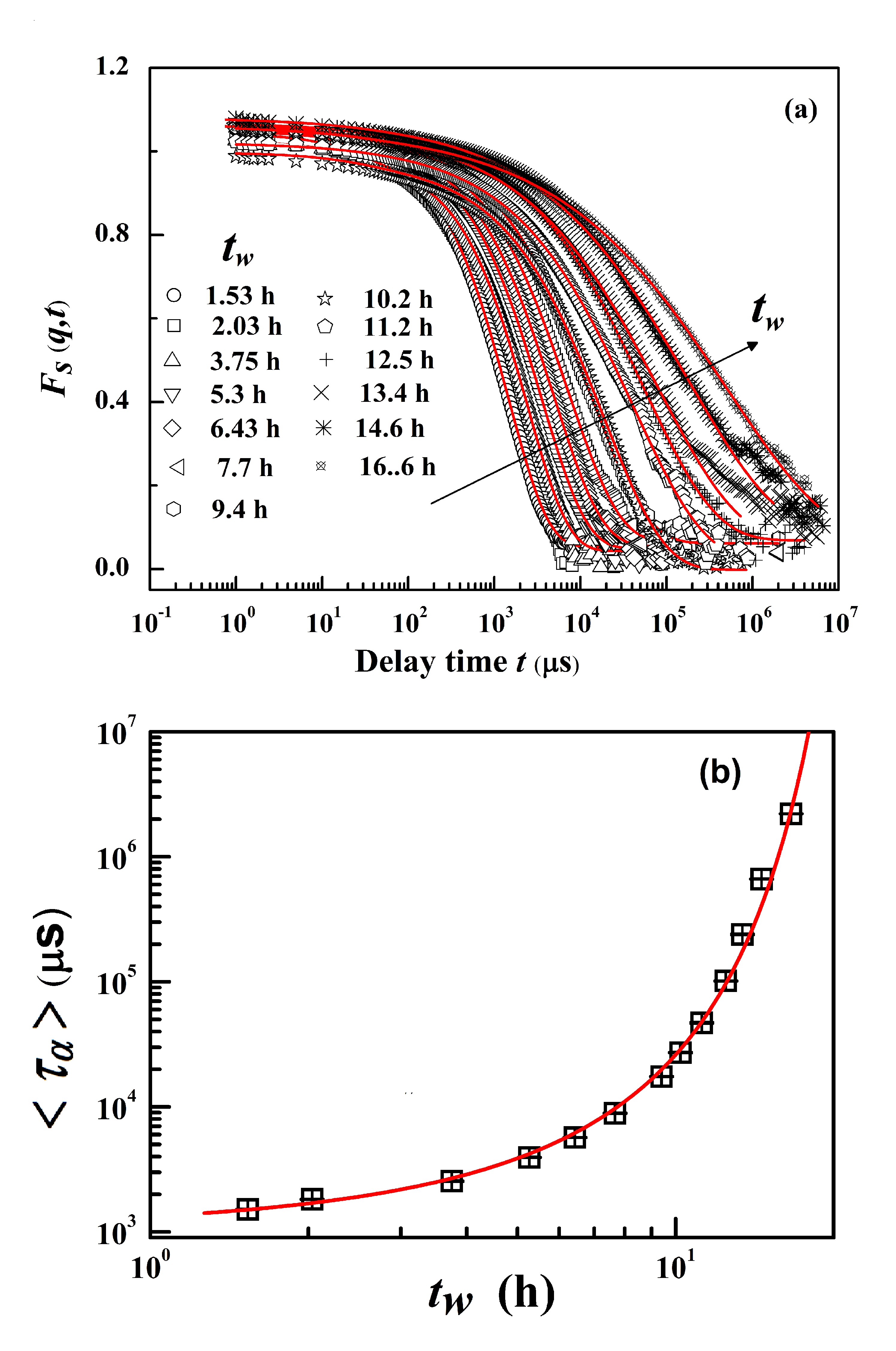}
\caption{(a) Self intermediate intensity scattering functions $F_{s}(q,t)$ decay curves vs. delay times recorded at various aging times ($t_{w}$) at $\theta$=90$^{\circ}$. Curves are shifted for better visibility. Solid lines are fits to the two step relaxation function given in Eq.3 and the fitting parameters obtained with increasing $t_{w}$ are given in the Supporting Information, Table S1. (b) Mean structural relaxation time ($<\tau_{\alpha}>$) as a function of $t_{w}$ for a Laponite-PS suspension.  The data plotted here is for a Laponite suspension of concentration $C_{L}$ = 3 wt.\%, and PS volume fraction $\phi_{PS}$ = 5.66 $\times$ 10$^{-5}$. The solid line is a fit to Eq.4.}  	
\label{VFT}
\end{figure}

The intensity auto correlation function $g^{(2)}(q,t)$ in DLS \cite{Bern} is related to the self-intermediate scattering function $ F_{s}(q,t) $ by the Siegert relation \citep{{pusey},{megen}},

\begin{equation}
 g^{(2)}(q,t)-1 = A|F_{s}(q,t)|^{2}
\label{eq2}
\end{equation}

where \begin{equation}
F_{s}(q,t)=\frac{1}{N}\left<\sum_{j}\exp\{iq.[r_{j}(t)-r_{j}(0)]\}\right>
\label{eq2}
\end{equation}

Here, $r_{j}(t)$ is the position of particle $j$ at time $t$, $A$ is the spatial coherence factor and the brackets represent an ensemble average. The $ F_{s}(q,t) $ or two-point correlation function quantifies the relaxation dynamics between delay times of 0 and $t$ seconds. $ F_{s}(q,t) $ {\it vs.} delay time plots are obtained using DLS experiments for aging Laponite clay suspensions at several aging times $t_{w}$. Representative plots are displayed in Fig.1(a). The normalized $ F_{s}(q,t) $ plots are seen to decay comparatively faster for smaller $t_{w}$ values. As $t_{w}$ increases, the decay of $ F_{s}(q,t) $ slows down considerably. Furthermore, the decay of $F_{s}(q,t)$ can be described as a two-step process \cite{saha}.  As reported earlier, the observed two-step relaxation function is a sum of a fast decaying exponential part and a comparatively slower stretched exponential part \cite{{saha},saha2,{ruzica1}} and can be expressed as    

\begin{equation}
F_{s}(q,t)= a\exp\{-t/\tau_{1}\}+(1-a)\exp\{-(t/\tau_{ww})^\beta\}  
\label{eq1}
\end{equation}

Here, $ \tau_{1} $ is the fast or secondary relaxation time (corresponding to the diffusion of a Laponite particle inside the cage formed by the neighbors), $\tau_{ww}$ is the  structural or primary $\alpha$-relaxation time (representing its cooperative diffusion to a neighbouring position), $\beta$ is a stretching exponent, and $a$ is the weight factor for the faster secondary relaxation process. The growth in the structural relaxation time $\tau_{ww}$ arises from an evolution of the screened interparticle electrostatic repulsion due to a gradual process of tactoid exfoliation \cite{ali_langmuir}. Fits of the experimental data to Eq.3 are shown in Fig.1(a) and the fitting parameters $a$, $\tau_{ww}$ and $\beta$, obtained from the fits for several $t_{w}$, are shown in Table S1 in Supporting Information (SI). $\tau_{1} = 30 \mu S$ is the fixed parameter in Eq.3 which corresponds to the diffusion of a single Laponite particle. The details of the calculation is presented in SI. The dependence of the mean structural relaxation time  $<\tau_{\alpha}>=(\tau_{ww}/\beta)(1/\Gamma(\beta))$ \cite{{LindsayJCP}} on $ t_{w}$ is obtained from the fits. The $<\tau_{\alpha}>$ is plotted in Fig.1(b) can be described by the Vogel-Fultcher-Tammann (VFT) law that governs  the dynamical slowdown process in fragile supercooled liquids \citep{{Angel},{Angel2}}, but with $ 1/T $ is replaced with $t_{w}$ \citep{{saha},{saha2}}. It is to be noted here that colloidal glasses have been studied successfully by invoking a VFT growth of the primary relaxation timescale, but with $1/T$ replaced by $\phi$ \citep{{Eric}}. Similarly for aging Laponite suspensions as investigated in earlier work \citep{saha,saha2,sahaLagm}, the VFT equation is as follows: 

\begin{equation}
<\tau_{\alpha}>=\tau_{o}\exp\left(\frac{Dt_{w}}{t_{\infty}-t_{w}}\right)
\label{eq2}
\end{equation}  

In this equation, $\tau_{o} = \tau_{\alpha}(t_w \rightarrow 0)$, $D$ is the fragility parameter (the inverse of $D$ quantifies the apparent deviation from the Arrhenius trend), and $ t_{\infty} $ is the waiting time at which the relaxation time diverges. For the data plotted in Fig.1(b), $\tau_{o}$ = 1060$\pm$ 65 $\mu$s, $D$ = 6.8 $\pm$  0.5 and $t_{\infty}$ = 31.3 $\pm$ 1.3 hours. Such a VFT like growth of $<\tau_{\alpha}>$  with increasing $t_{w}$ as the Laponite suspension approaches eventual kinetic arrest is strongly reminiscent of the observations reported in fragile molecular supercooled liquids with decreasing temperature $T$ and in many hard sphere colloidal glasses with increasing volume fraction $\phi$. 
	
	Typically, the fragile behavior of glass-forming liquids as they approach the glass transition is rationalized by enumerating the number of particles engaged in slow correlated motion, $N_{corr}$, in the DHs and the growing sizes of these regions of cooperative motion. Moreover, the $\beta$ values, obtained from Eq.3 are plotted as a function of $t_{w}$ and shown in Fig.2(a). The decrease in $\beta$ implies the width of time scale distributions characterising the dynamics of independently relaxing DHs increases monotonically with increasing $t_{w}$.  The correlated dynamics can therefore be modelled by assuming linear superposition of exponential relaxation of DHs, with each DH having its own relaxation time $\tau$ \cite{{Dalle}}. Distribution functions of relaxation times ($ G_{ww} (\tau) $) are calculated for the Laponite system using the following Kohlrausch-Williams-Watts equation \citep{LindsayJCP} as $t_{w}$ increases 
 
\begin{equation}
\rho_{ww}(\tau)=-\frac{\tau_{ww}}{\pi \tau^{2}}\sum_{k}^{\infty}\frac{(-1)^k}{k!}\sin(\pi \beta k)\Gamma(\beta k+1) (\frac{\tau}{\tau_{ww}})^{(\beta k+1)}
\label{eq2}
\end{equation}

The distribution $ G_{ww} (\tau)= \tau \rho_{ww}(\tau) $ at four different $ t_{w} = 1.53, 6.43, 10.2 $ and $16.6$ h  are plotted in Fig.2(b). A Laponite  suspension lying in the liquid-like regime ({\it i.e.} at small $ t_{w} $ values) is characterised by a $ G_{ww} $ which shows a sharp peak at $ \tau/\tau_{ww}$ close to 1. In contrast, as the sample ages towards a non-ergodic state, a broad distribution of $ G_{ww} $ having peak position at $ \tau / \tau_{ww} > 1 $ is observed. This suggests that at small sample ages, the relaxation timescales characterising the reorganization dynamics is more likely to have values within a narrow range. The broad distribution at longer $ t_{w} $  can be rationalized by considering the presence of simultaneous fast and slow moving cooperatively rearranging regions having their own independent relaxation times. The broadening of $ G_{ww} $ with increase in the control parameter $t_w$ is reminiscent of observations in fragile supercooled liquids with decreasing $T$ \cite{{Ediger2},{Stillinger}} and strongly indicates the presence of heterogeneous dynamics in aging colloidal Laponite suspensions.

\begin{figure}[!t]
\includegraphics[width=3.3in]{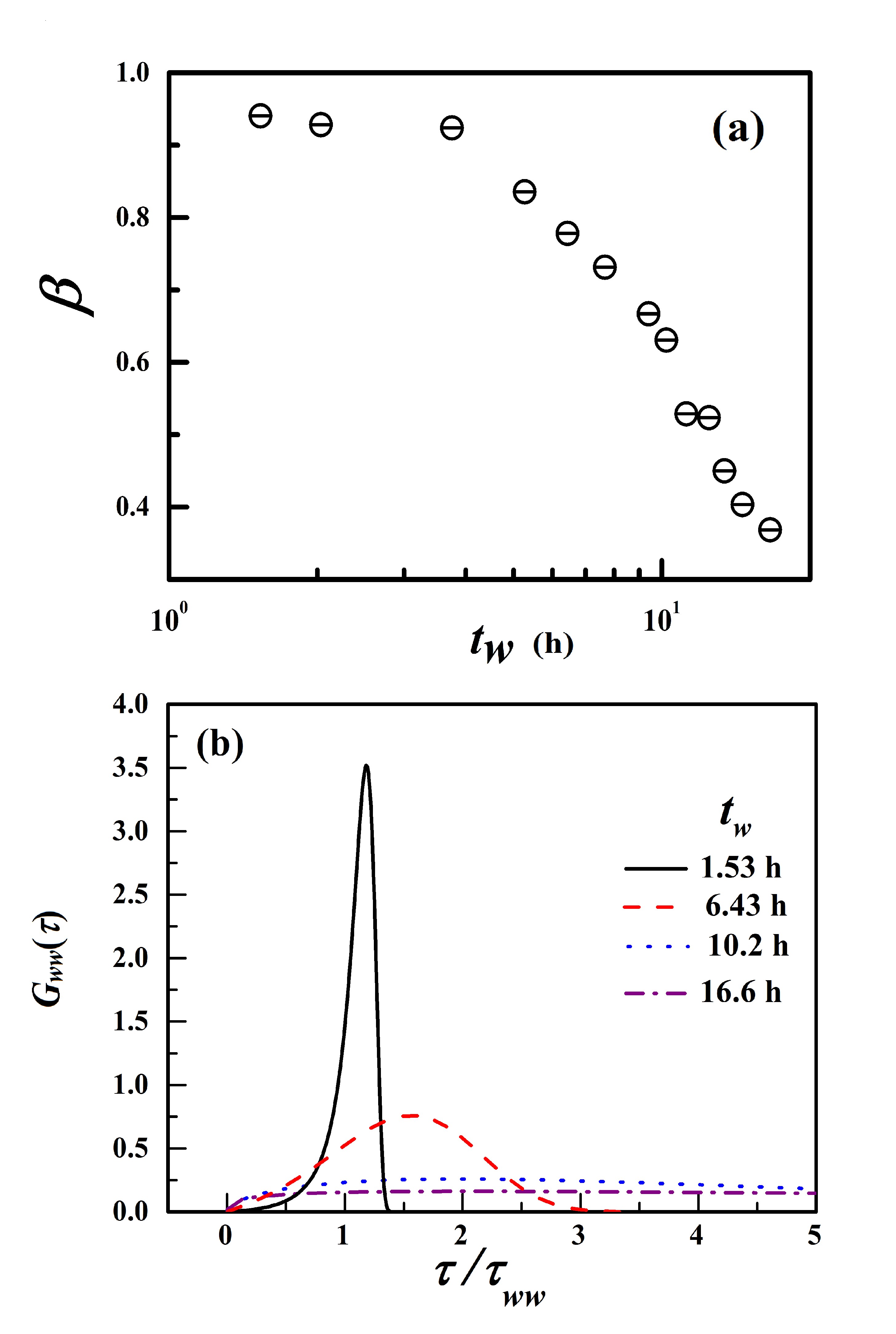}
\caption{(a) The stretching exponent $ \beta $ vs.$t_{w}$. (b) Distribution of relaxation time scales $ G_{ww} (\tau) $ plotted at various $ t_{w} $ for the Laponite-PS system of Fig.1}
\label{}
\end{figure}

\subsection{Characterizing dynamical heterogeneities by evaluating three-point correlation functions}
   
Characterizing $N_{corr}$ of DHs in experiments is a challenging task. Theoretically, information about the sizes of the DHs and $ N_{corr} $ are embedded in the four point dynamic susceptibility which takes into account correlations in both space and time. A detailed theoretical and mathematical treatment of the four point correlation function and its relation to $N_{corr}$ can be found in the literature \citep{{Lacevic},{Toninelli}}. The four-point dynamic susceptibility $\chi_{4}(q,t)$ is related to the fluctuating part of the self-intermediate scattering function $\delta F_{s}(q,t)$ by the following equation:  
	
\begin{equation}
\chi_{4}(q,t)=N_{corr}<\delta F_{s}(q,t)^{2}>
\label{eq2}
\end{equation}  

Experimentally therefore, $\chi_{4}$ can be obtained by resolving the dynamical behavior of the observable in both space and time. In contrast to molecular systems, colloidal particles have comparatively larger sizes (10 $n$m - 10 $\mu$m). This feature, and the easy tunability of inter-particle interactions, make colloidal systems robust model candidates for experimental and simulation studies of a variety of physical phenomena. Indeed, $\chi_{4}$ has been extracted using advanced microscopy techniques for colloidal glasses and in simulations \citep{{Berthier},{Yodh},{Durian}}. However, since computation of spatial correlations from experimental data on molecular and nanocolloidal suspensions, {\it eg.}  Laponite, is extremely difficult, the four point dynamic susceptibility calculation method invariably fails in these cases. In order to study DHs in the aforementioned systems, therefore, the three-point correlation function, which is the lower bound of the four-point correlation function, has been introduced \citep{{Berthier2},{Berthi},{Berthi2}}. It can be accessed experimentally by probing the sensitivity of the two-point correlation function (dielectric response or the scattering function) to external control parameters such as $T$ in supercooled liquids and $\phi$ in colloidal systems. For hard sphere colloids, $ \chi_{4} (q,t) $ is written as \citep{{Berthier2},{Brambilla}}
  
 \begin{equation}
\chi_{4}(q,t)=\chi_{4}(q,t)|_{\phi}+\rho k _{B} T k _{T}[\phi\chi_{\phi}(q,t)]^{2} 
\label{eq2}
\end{equation}
 where $\chi_{4}(q,t)|_{\phi}$ is the fixed density value, $\rho$ is the particle number density, $k _{T}$ is the isothermal compressibility and $\chi_{\phi}(q,t)$ is the derivative of $F_{s}(\tau_{\alpha},\beta)$ with respect to $\phi$. The second term in Eq.7 is the three-point dynamic susceptibility and can be accessed experimentally. The derivative of $F_{s}(\tau_{\alpha},\beta)$ with respect to $\phi$, $\chi_{\phi}(q,t)$, has been obtained in dynamic light scattering experiments for colloidal systems \citep{{Brambilla}}, while in supercooled liquids, the thermal derivative of the dielectric response $ \chi_{T} $ has been used to study $N_{corr}$ \cite{Dalle}.
 In these experiments, $\chi_{\phi}(q,t)$ and $ \chi_{T} $ were evaluated by probing, respectively, the scattering function and dielectric response at infinitesimal regular intervals of $ \phi $ and $T$. The $\phi$-dependence of $F_{s}(q,t)$ or the $T$ dependence of the dielectric spectrum are fitted with polynomial functions, with the derivatives of the fitted curve with respect to the control parameter yielding the three-point susceptibility. The use of the three point correlation function formalism to calculate $N_{corr}$ in a wide range of glass forming liquids using dielectric and light scattering data has been successfully demonstrated by  Dalle-Ferrier {\it et al.} \citep{{Dalle},{Dalle2}}. The authors showed that $\chi_{\phi}(q,t)$ and $\chi_{T}$ show peaks similar to $\chi_{4}$, with the peak height being proportional to $N_{corr}$.

\begin{figure}[!t]
\includegraphics [width=3.3in]{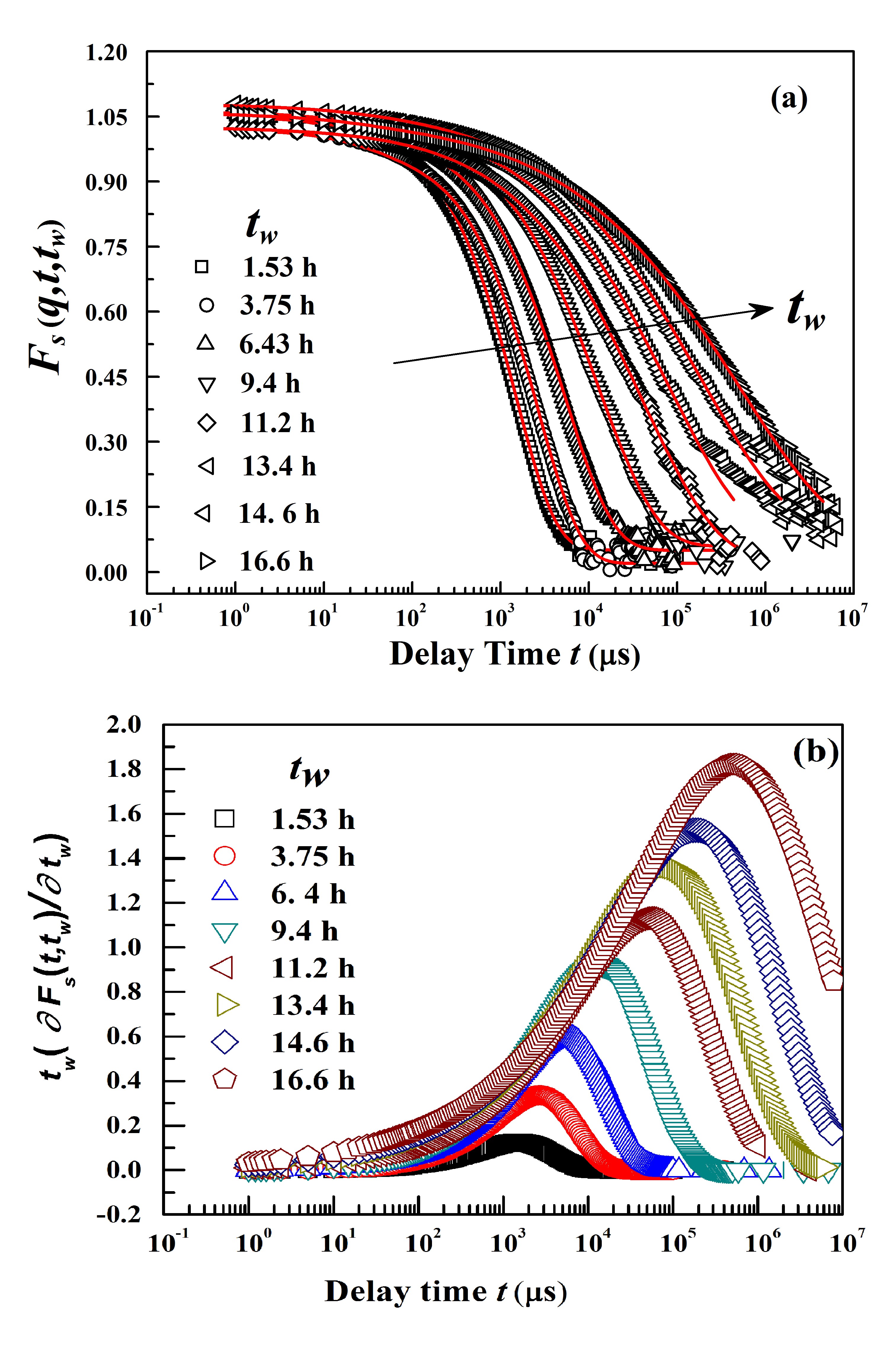}

\caption{(a)$ F_{s}(q,t,t_{w}) $ decay curves at various $ t_{w} $ for Laponite-PS suspensions (Laponite concentration $C_{L}$ = 3 wt.\%, and PS volume fraction $\phi_{PS}$ = 5.66 $\times$ 10$^{-5}$). The solid lines are fits to Eq.8. (b) Plots of $\chi_{t_{w}}$ \textit{(q,t)} = $ t_{w} \partial F_{s}(q,t,t_{w})/\partial t_{w} $ as a function of delay time $t$ for the Laponite-PS sample (Fig.1)}  	
\label{}
\end{figure}

\begin{figure}[!t]
\includegraphics [width=3.3in]{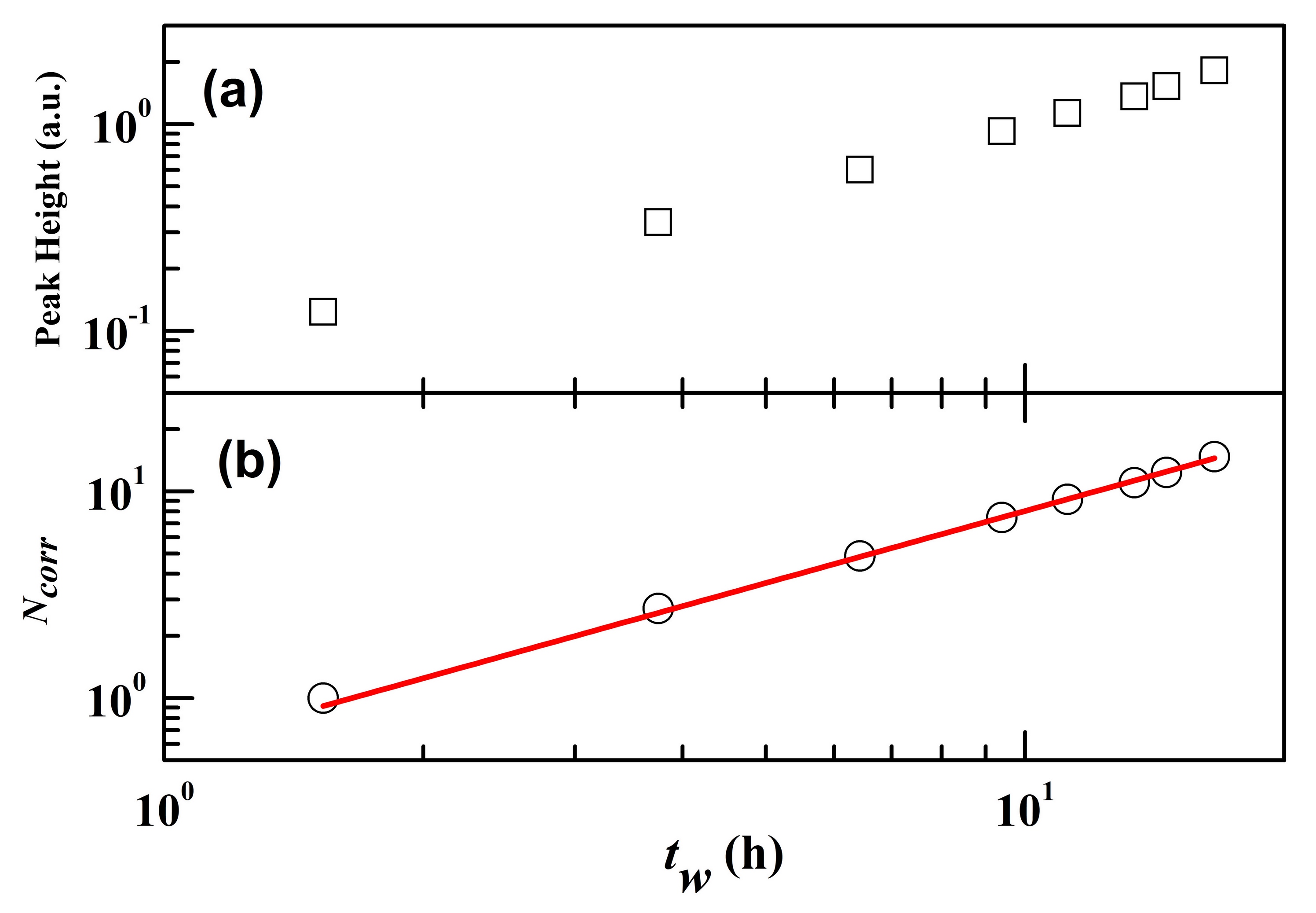}
\caption{(a) Growth of peak height and (b) $N_{corr}$ calculated from the data in Fig.3(b) as $t_{w}$ increases towards the non-ergodic state. The solid line in (b) is a power law fit of the form $N_{corr} = B (t_{w})^{\gamma}$ where $\gamma$ = 1.17 $\pm$ 0.02 and $B$ = 0.56 $\pm$ 0.04.}  	
\label{}
\end{figure}

\begin{figure}[!t]
\includegraphics [width=3.3in]{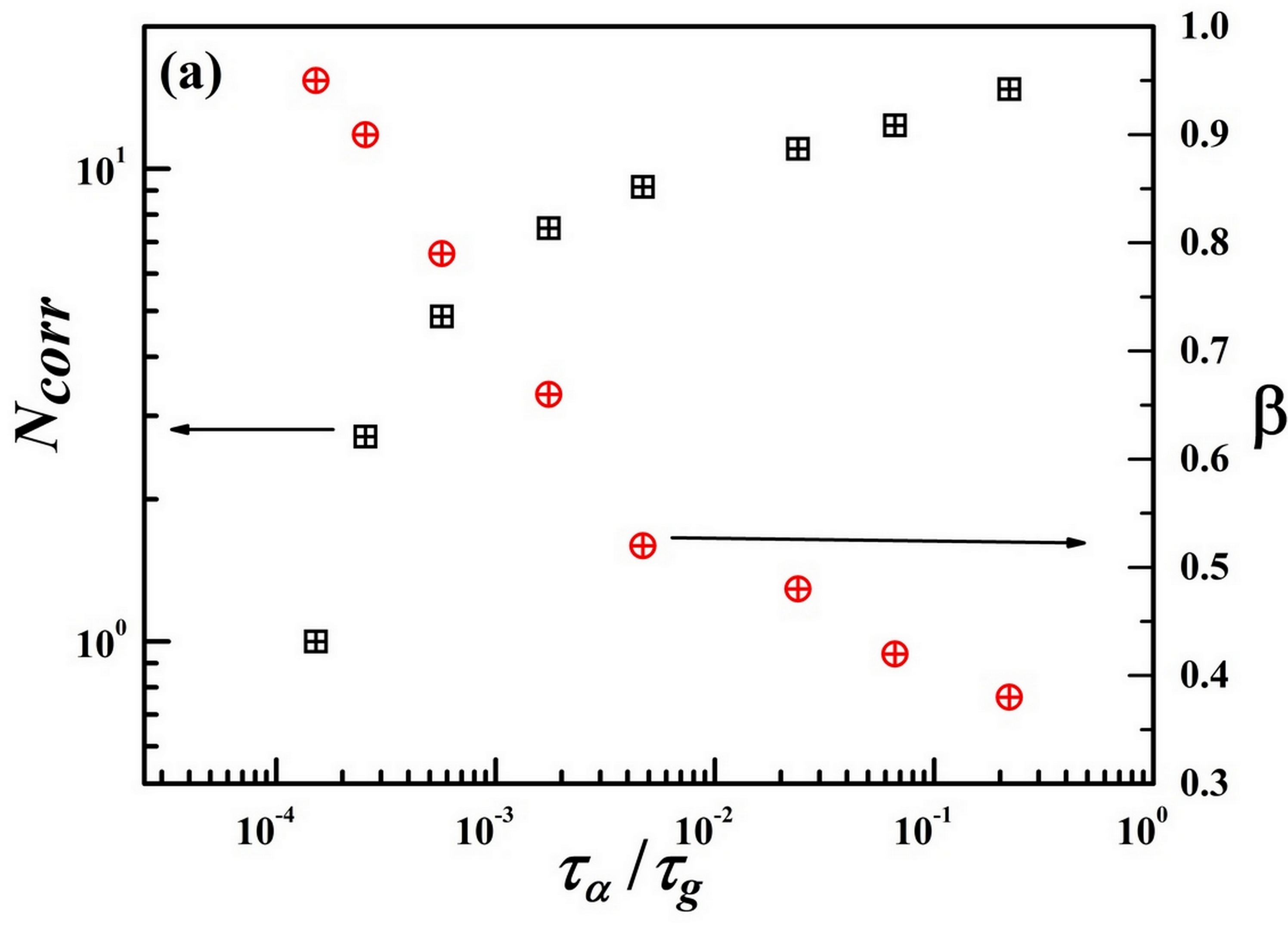}
\caption{(a) Plot of $N_{corr}$ (black symbols) and $\beta$ (red symbols) {\it vs.} $\tau_{\alpha}/\tau_{g}$. $N_{corr}$ shows a monotonic increase, with the initial rapid increase slowing down considerably at high $\tau_{\alpha}/\tau_{g}$ while $ \beta $ shows a monotonic decrease.}	
\label{}
\end{figure}

In our study, the procedure outlined above is implemented to obtain $N_{corr}$ for aging Laponite colloidal suspensions after replacing $\phi$ with $t_{w}$. The dependence of $F_{s}(q,t)$ on $t_{w}$ is shown in Fig.3(a). It has been shown earlier that the decay curves in Fig.3(a) fit well to Eq.3 \cite{saha,saha2}. In order to obtain derivatives of $F_{s}(q,t)$ with respect to $t_{w}$, Eq.3 is parametrized with Eq.4 to obtain

\begin{equation}
\begin{split}
F_{s}(q,t,t_{w})= a\exp\{-t/\tau_{1}\} + \\
(1-a)\exp\left\{-\left(t/\left\{\tau_{o}\exp\left(\frac{Dt_{w}}{t_{\infty}-t_{w}}\right)\right\}\right)^\beta\right\}
\end{split}
\label{eq3}
\end{equation}  

The $F_{s}(q,t,t_{w})$ decay curves, fitted to Eq.8, are depicted in Fig.3(a) using solid lines. The fitting parameters thus obtained  are in good agreement with the parameters extracted by fitting the data to Eq.3. The fitting parameters are tabulated in Table S2 of the SI.  
The three-point correlation function for aging Laponite system, $\chi_{t_{w}}(q,t)$, is obtained by differentiating the fitted curves shown in Fig.3(a) with respect to $t_{w}$, {\it i.e.} $\dfrac{\partial F_{s}(q,t,t_{w})}{\partial t_{w}} $. The calculated derivatives, shown in Fig.3(b), exhibit peaks, with the peak heights increasing and the peak positions shifting to higher delay times $t$ with increasing $t_{w}$. The observed growth of the peak height of the $\chi_{t_{w}}(q,t)$, obtained using $t_{w}$ as the relevant control parameter, demonstrates the growing dynamical heterogeneities and the monotonically increasing $N_{corr}$ during the spontaneous aging process of Laponite colloidal suspensions. The height of the peak, plotted in Fig.4(a) as a function of $t_{w}$, is proportional to $N_{corr}$ \cite{Dalle}. Next, $N_{corr}   \propto  [t_{w}(\dfrac{\partial F_{s}(q,t,t_{w})}{\partial t_{w}})]$ is calculated with the assumption that for the smallest waiting times probed in this work (very low $t_{w}$), correlated particle motion can be ruled out completely (with particles moving independently; $N_{corr}=$ 1). The proportionality pre-factors in this relation for $N_{corr}$ can be assumed to be independent of $t_{w}$. Finally, $[t_{w}(\dfrac{\partial F_{s}(q,t,t_{w})}{\partial t_{w}})]$ is normalized with respect to the earliest $t_{w}$ value at all $t_{w}$ to obtain $N_{corr}$ as a function of $t_{w}$. The corresponding $ N_{corr} $ plot is given in Fig.4(b). If $\tau_{\alpha}=10$ s is defined as the non-ergodic transition point \citep{Dalle}, a systematic growth of $N_{corr}$ with increasing $t_{w}$ is observed. The growth of $N_{corr}$ shows a power law dependence on $t_{w}$ ($N_{corr} = B (t_{w})^{\gamma}$, where the exponent $\gamma$ = 1.17 $\pm$ 0.02, $B$ = 0.56 $\pm$ 0.04) and is shown by a solid line in Fig.4(b). Similar power law growth of $N_{corr}$ is observed for the measurements (upto $t_{w}=9.6h$) taken at a lower scattering angle $\theta$ = 60$^{\circ}$ (thereby probing longer length scales) and the plot is given in the SI in Fig. S1.

In Fig.5, $N_{corr}$ is plotted as a function of $\tau_{\alpha}$ normalized with $\tau_{g}=10$ s to observe its evolution. $\beta$ values are also depicted in the same figure to correlate the evolution of the time scale distributions with growth of $N_{corr}$. The decrease of $\beta$ (or increase of the width of distribution of relaxation time scales, $ G_{ww} $, as shown in Fig.2(b)) as Laponite suspensions age towards the glass transition indicates that the dynamics turn progressively heterogeneous.This implies growing correlations in the system {\it i.e.} the appearance of a large number of regions of correlated groups of particles with their own relaxation times, as the suspension ages. This can be understood from the evolution of $N_{corr}$ as a function of $\tau_{\alpha}/\tau_{g}$ (shown in Fig.5). The appearance of dynamic heterogeneities of larger sizes (characterised by $N_{corr}$) results in a slowdown of the dynamics (as reported in Fig.1(b)) which manifests as a fragile supercooled liquid-like growth of $<\tau_{\alpha}>$ (VFT-like plot in Fig.1(b)). However, it is seen that $N_{corr}$ grows rapidly at small $\tau_{\alpha}$, but slows down considerably close to the non-ergodic transition point. Remarkably, this trend is in close agreement with an observation in supercooled liquids, where  $N_{corr} $ is reported to show a power-law dependence at large $T$ ({\it i.e.} in the liquid regime) and a logarithmic growth at low $T$ ({\it i.e.} near the glass transition) \citep{{Dalle},{Dalle2}}. This slow increase of $N_{corr}$ close to the non-ergodic transition is attributed to activated dynamics. In this regime, cooperative motion becomes increasingly difficult as the structural rearrangement time slows down dramatically as a consequence of the very high viscosity of the system. This results in large $<\tau_{\alpha}>$ values. The increasing values of $ N_{corr} $ with $t_{w}$, estimated from the three-point dynamic susceptibility in the present study, provides a quantitative measure of the fragile supercooled liquid-like heterogeneous dynamics in aging colloidal Laponite suspensions. 
 
\section{Conclusions}
Dynamical heterogeneities in aging Laponite colloidal suspensions are investigated by the three-point dynamic susceptibility formalism with respect to a new control parameter, {\it viz.} $t_{w}$. Laponite colloidal suspensions exhibit fragile supercooled liquid-like dynamics, with the average primary relaxation time, $<\tau_{\alpha}>$, in its characteristic two-step relaxation process, exhibiting VFT growth as a function of $t_{w}$. Three-point dynamic susceptibilities are computed from DLS experiments by taking the derivative of the two-point scattering decay function, $F_{s}(q,t)$, with respect to  $t_{w}$. The three-point dynamic susceptibility exhibits a peak, with the peak height growing with increasing $t_{w}$. Our calculations show that $N_{corr}$, the number of particles participating in correlated motion thereby setting the size of the dynamical heterogeneity, shows a power law increase with increase in  the aging time $t_{w}$ of Laponite suspensions. Furthermore, we show that the growth of $N_{corr}$ is initially quite fast, before  slowing down close to the glass transition.  We believe that our study provides valuable insight into the approach of Laponite suspensions towards kinetic arrest by demonstrating that  the growth of correlations in aging Laponite suspensions closely resemble the observations reported for fragile supercooled liquids\cite{{Dalle}}. Unlike in fragile supercooled liquids, however, the dynamics of Laponite suspensions are driven by athermal processes such as long-range screened inter-particle electrostatic repulsions and tactoid exfoliation. Given the similarities in the dynamical slowdown processes and the growth of correlations in Laponite suspensions and fragile supercooled liquids, the ubiquitous kinetic slowdown in glass-forming liquids, driven by a growth in heterogeneous dynamics, could well be a universal feature of fragile glass-formers. 

%
\end{document}